\documentclass[iop]{emulateapj}

\begin{document} 
\submitted{Accepted for publication in RNAAS}

\title{Gravity-darkening exponents for MESA stellar evolution models}

\email{Corresponding author: gtorres@cfa.harvard.edu}

\author{A.\ Claret}
\affiliation{Instituto de Astrof\'{\i}sica de Andaluc\'{\i}a, CSIC, Apartado 3004, 18080 Granada, Spain}
\affiliation{Dept.\ F\'{\i}sica Te\'{o}rica y del Cosmos, Universidad
  de Granada, Campus de Fuentenueva s/n, 10871 Granada, Spain}

\author{G.\ Torres}
\affiliation{Center for Astrophysics $\vert$ Harvard \& Smithsonian,
  60 Garden St., Cambridge MA 02148, USA}

\begin{abstract}
In this Research Note we present new gravity-darkening exponents
($\beta$) for several stellar evolution models from the ZAMS up to the
giant phase. The models were computed using the MESA code (version
7385) for the composition $X = 0.70$ and $Z = 0.02$, adopting A09
opacities and a mixing length parameter of $\alpha_{\rm MLT} = 1.84$.
Results were calculated using the triangle strategy, for initial stellar masses of
1.0, 1.5, 2.0, 2.5, 3.0, 5.0, 7.0, 10.0, 15.0, and 20.0~$M_{\sun}$,
and provide an update on previously available $\beta$ values that
used older models.
\end{abstract}


\section{Introduction}

In a pioneering paper, \cite{vonZeipel:1924} quantified the effect of
gravity-darkening in stars by establishing the relationship between the local
surface gravity, $g$, and the effective temperature, $T_{\rm eff}$,
for the case of envelopes in radiative equilibrium. The corresponding
equation can be written more generally as:
\begin{eqnarray}
 {\bf F} = -{4 a c T^{3}\over{3 \kappa \rho}}{dT\over{d\Phi}} g^{\beta}~,
\end{eqnarray}
or equivalently 
\begin{eqnarray}
 T_{\rm eff}^4 \propto g^{\beta}~.
\end{eqnarray}
In the first of these expressions, $\mathbf{F}$ is the vector of
radiative energy flux, $\Phi$ is the total potential,
$T$ is the local temperature, $\kappa$ the opacity, $\rho$ the local
density, $a$ the radiation pressure constant, and $c$ is the speed of
light in vacuum. The gravity-darkening exponent (GDE)
$\beta$ is a bolometric quantity, and is equal to unity for envelopes in
radiative equilibrium.
Decades later, \cite{Lucy:1967} addressed the case of envelopes in
convective equilibrium, and derived a single average value of $\beta
\approx 0.32$. 

Some 30 years later, \cite{Claret:1998} revisited the
issue of the calculation of $\beta$, and introduced the so-called
method of triangles for its derivation \citep[see][]{Kippenhahn:1967},
valid for models over the entire H-R diagram, including those for stars
with convective envelopes.
This efficient and expedient method allows for the consideration
of evolutionary changes in both the stellar interior and the outer
layers. It provides continuity for the theoretical values of the GDE 
across the
H-R diagram, thus avoiding the previous physically unrealistic situation
characterized by a step function in $\beta$ between radiative and
convective envelopes (1.0--0.32). Here we use this method as well.

Empirical determinations of $\beta$ from W~UMa eclipsing binaries,
which experience strong proximity effects leading to significant
gravity darkening, generally support the continuity of the GDE across the
convective/radiative boundary, as predicted by theory
\citep[see, e.g., Fig.~3 by][]{Claret:2003}. While some of those early
estimates have suffered from biases caused by limitations in the photometry,
much improved determinations should now be possible with
modern instrumentation and the availability of high-precision and
high-cadence light curves from space-based facilities, such as NASA's
Kepler/K2 and TESS missions, and others.

GDEs have been used extensively in the interpretation of light curves
of eclipsing binaries, enabling the derivation of accurate stellar
properties (component masses, radii, temperatures, etc.). More recently,
they have also found use in the transiting exoplanet field,
particularly for the study of rapidly rotating host stars.
The precision of the light curves now available
places more stringent demands on the accuracy of stellar parameters
adopted from theory, including $\beta$.
We expect the new values of the GDEs presented in this Note to be helpful
to the community for studies such as those mentioned above.
The triangle strategy applied here works well for this purpose, within the theoretical
uncertainties typical of stellar evolution theory, and has
a number of advantages:

\noindent $\bullet$ It is a self-consistent method, as the same
evolutionary models are used to compute continuous $\beta$
values as well as many other stellar properties. These include the stellar radius,
$T_{\rm eff}$, internal pressure and other thermodynamic variables,
apsidal-motion constants, the moment of inertia, the gravitational
potential energy, and additional parameters needed for the calculation
of timescales of orbit circularization and spin-orbit synchronization
in binary systems, among others;

\noindent $\bullet$ The changes in the chemical profile in the envelope and
in the interior are taken into account at each timestep;

\noindent $\bullet$ The method does not present possible effects of
sphericity, given the spherical symmetry of the structure equations;

\noindent $\bullet$ It can be adapted to any stellar evolution code;

\noindent $\bullet$ It allows the study of the influence of optical depth
on the GDE calculations.

In this Note we applied the method to the calculation of $\beta$ with
the widely used Modules for Experiments in Stellar Astrophysics code
\citep[MESA;][and references therein]{Paxton:2019} over a wide range
of stellar masses. The use of a state-of-the-art code such as MESA to
generate stellar evolution models provides a needed update to the
calculations previously available \citep{Claret:1998},
as many of the key physical ingredients have changed over the years.

\section{Results}

Our calculations used MESA version 7385, for consistency with our
earlier work presenting calculations of apsidal motion constants
obtained with the same code,
for use in the study of eclipsing binaries \citep[see][]{Claret:2019,
Claret:2021, Claret:2023, Claret:2024}.
Here we computed
$\beta$ values for evolutionary models from the zero-age main sequence (ZAMS) up to the
giant phase, and initial masses of 1.0, 1.5, 2.0, 2.5, 3.0, 5.0,
7.0, 10.0, 15.0, and 20.0~$M_{\sun}$ (see the machine readable version of Table 1). Results are presented for
the composition $X = 0.70$ and $Z = 0.02$, with `A09' radiative opacities from
\cite{Asplund:2009}, and an adopted solar-calibrated mixing-length parameter
$\alpha_{\rm MLT} = 1.84$.
The evolutionary tracks provide $\beta$ along with other customary
quantities ($\log L$, $\log g$, $\log T_{\rm eff}$, and current mass)
as a function of age, and are available electronically with this Note.
The first few lines of the table containing all 10 evolutionary tracks
is shown in Table~\ref{tab:tab1}, to illustrate the content.

\setlength{\tabcolsep}{4pt}
\begin{deluxetable}{cccccc}
\tablewidth{0pc}
\tablecaption{Merged Evolutionary Tracks \label{tab:tab1}}
\tablehead{
\colhead{Age} &
\colhead{$\log L$} &
\colhead{$\log g$} &
\colhead{$\log T_{\rm eff}$} &
\colhead{Mass} &
\colhead{$\beta$}
\\
\colhead{(yr)} &
\colhead{($L_{\sun}$)} &
\colhead{(cgs)} &
\colhead{(K)} &
\colhead{($M_{\sun}$)} &
\colhead{}
}
\startdata
1.00000D+05  &  $-$0.16450  &  4.53493  &  3.74476  &  1.00000  &  0.40517 \\
2.20000D+05  &  $-$0.17110  &  4.53800  &  3.74388  &  1.00000  &  0.40726 \\
3.64000D+05  &  $-$0.17814  &  4.54076  &  3.74281  &  0.99999  &  0.40853 \\
5.36800D+05  &  $-$0.18335  &  4.54336  &  3.74216  &  0.99999  &  0.40911 \\
7.44160D+05  &  $-$0.18829  &  4.54579  &  3.74153  &  0.99999  &  0.40984 
\enddata
\tablecomments{Table 1 is published in its entirety in the electronic 
edition of the {\it RNAAS}.  A portion is shown here 
for guidance regarding its form and content. The full version contains the data for all 10 initial masses from this work.}
\end{deluxetable}

\acknowledgments
{The Spanish MINC/AEI (grants PID2022-137241NB-C43 and
PID2019-107061GB-C64) are gratefully acknowledged for their support
during the development of this work.  We also acknowledge financial
support from grant CEX2021-001131-S, funded by MINC/AEI
10.13039/501100011033.  This research has made use of NASA's
Astrophysics Data System Abstract Service.}

\end{document}